\documentclass[prb,aps,twocolumn,floatfix]{revtex4}

\usepackage{dcolumn}
\usepackage{amssymb}
\usepackage{graphicx}

\newcommand{\be}{\begin{equation}}
\newcommand{\bea}{\begin{eqnarray}}

\newcommand{\diff}{{\rm d}}
\newcommand{\ee}{\end{equation}}
\newcommand{\eea}{\end{eqnarray}}

\begin{document}

\title{From laser cooling to aging: a unified L\'evy flight description}
\author{Eric Bertin$^{1,2}$ and Fran\c{c}ois Bardou$^{3,}$}
\thanks{Deceased.}
\affiliation{$^1$Universit\'e de Lyon, Laboratoire de Physique, ENS Lyon,
CNRS, 46 All\'ee d'Italie, F-69007 Lyon, France\\
$^2$Department of Theoretical Physics, University of Geneva, CH-1211 Geneva 4, Switzerland\\
$^3$IPCMS, 23 rue du Loess, BP 43, F-67034 Strasbourg Cedex 2, France}


\begin{abstract}
Intriguing phenomena such as subrecoil laser cooling of atoms, or aging
phenomenon in glasses, have in common that the systems considered do not
reach a steady-state during the experiments, although the experimental
time scales are very large compared to the microscopic ones.
We revisit some standard models describing these phenomena, and reformulate
them in a unified framework in terms of lifetimes of the microscopic states
of the system.
A universal dynamical mechanism emerges, leading to a generic
time-dependent distribution of lifetimes, independently of the physical
situation considered.
\end{abstract}

\maketitle

\section{Introduction}

One of the important paradigmatic changes in the twentieth century physics
was the introduction of probabilities, and hence fluctuations, as a
central concept in the description of the physical world.
This arose through the development of both quantum mechanics
and statistical mechanics.
Along this line of thought, a physical observable is often decomposed
into an average value plus some fluctuations.
Accordingly, fluctuations are most often seen as small perturbations
around a well-defined mean value, meaning that the probability distribution
of the observable is narrow.

However, it has become clear since the mathematical work of Paul L\'evy
in the 1930's,\cite{Levy,Feller} that interesting situations arise
when the probability
distribution becomes so broad that the average value of the
observable is not even defined (i.e., it is formally infinite).
This appears in particular when the tail of the distribution $p(x)$
of the variable $x$ behaves as a power law
\be
P(x) \sim \frac{c}{x^{1+\alpha}}, \qquad x \to \infty,
\ee
with $0<\alpha<1$.
In the early 1970's, it was realized that L\'evy's
results could be relevant to the study of random walks,\cite{Mon73,Shl74}
with physical applications for instance in
disordered conductors.\cite{Mon75}
If the broadly distributed variable corresponds to the sojourn times
in some microscopic states of a system, some interesting and experimentally
testable properties arise, like anomalous diffusion, or aging.
In the last fifteen years, this topic has become more and more popular in the
physics research community. Broad distributions have found applications
in physical processes as diverse as turbulent and chaotic
transport,\cite{SZK93,SWS93,Min96}
glassy dynamics,\cite{Shl88,BG,Bouchaud} random walks in
solutions of micelles,\cite{OBL90} diffusion of spectral lines in
disordered solids,\cite{ZuK94,Barkai}
fluorescence of single nanocrystals,\cite{Brokmann}
laser trapped ions,\cite{MEZ96,KSW97}
or laser cooling of atomic gases\cite{BBE94,BBA02}
(for which the Nobel Prize was awarded in 1997), to quote only some of them.

The aim of this paper is to present a brief overview, within a unified
formalism, of some elementary physical models in which the dynamics
involves broad distributions of sojourn times, and to highlight the
surprising phenomena that may appear from such a simple dynamics.
Using simple probabilistic arguments, we show that the equilibrium
distribution becomes non normalizable in a given parameter range,
thus ruling out the existence of an equilibrium state.
We then describe the dynamics of the different models in terms of
lifetime variables --the mean time spent in a given microscopic state.
This leads to a universal characterization of systems with
non-stationary dynamics at large time. In this description, the common
feature of these systems is the presence of power-law tails in the
distribution of lifetimes of the microscopic states,
leading to an infinite average lifetime.
We also compute the dynamical probability of lifetimes, that is, the
probability for the system to occupy at time $t$ a state with a given
lifetime, and find at large time a universal form for this distribution.

The paper is organized as follows. In Sect.~\ref{sect-models},
we give two examples of physical situations where the time spent by the
system in different microscopic states becomes broadly distributed, and
present simple models aimed at describing these physical situations.
In Sect.~\ref{sect-life-time}, we show that the two models
can be described within a unified framework, and compute the corresponding
dynamical distribution of probability, before discussing some physical
phenomena that can be explained by these results.
Finally, some technical details are reported in Appendices \ref{apME}
and \ref{apA}, and we suggest in Appendix \ref{sugg-pb} a project
appropriate for students.

\section{Simple physical examples} \label{sect-models}

\subsection{Laser cooling of atoms}

Laser cooling of atomic gases consists in reducing the momentum spread of
atoms thanks to momentum exchanges between atoms and photons.
The so-called ``subrecoil'' laser cooling consists in reducing
the momentum spread of the atoms
to less than a single photon momentum $\hbar k$,
where $k=2\pi/\lambda$ is the wavevector of the light, and $\hbar$
is the reduced Planck constant.
This cooling is achieved by introducing a momentum dependence in the photon
scattering rate, in such a way that it decreases strongly
in the vicinity of $p=0$, where $p$ is the magnitude of the atomic
momentum $\mathbf{p}$. When an atom reaches by chance this region of
momentum space, it tends to stay there for a long time,
since the photon scattering rate is very low.
This leads to an accumulation of atoms at small momenta, that is, to 
a cooling of the atoms.\cite{BBE94,BBA02}

\begin{figure}[t]
\centering\includegraphics[width=7.5cm,clip]{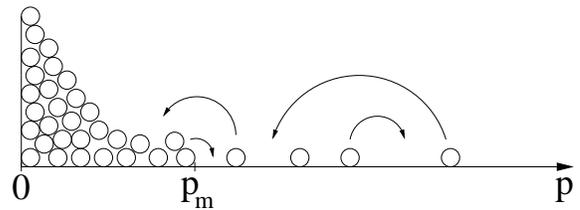}
\caption{\sl Schematic picture of the subrecoil laser cooling process
in momentum space.
Photon scatterings induce a random walk for the momentum $p$ of the atoms
($p$ is the magnitude of $\mathbf{p}$), with steps of the order of
$\hbar k$, the photon momentum.
The mean sojourn time $\tau(p)$ at momentum $p$ becomes very large
for small momenta (i.e.~for $p \ll p_m$), thus generating
an accumulation of atoms in the vicinity of $p=0$.
}
\label{fig-laser}
\end{figure}

The mechanism of subrecoil laser cooling is illustrated
in Fig.~\ref{fig-laser}.
Anytime a photon is absorbed and spontaneously reemitted by an atom,
the atomic momentum undergoes a random jump of the order of $\hbar k$,
since spontaneous emission occurs in a random direction.
Thus, the repetition of absorption-spontaneous emission cycles generates
a random walk of the momentum of the atom,
with momentum-dependent time intervals between two jumps.
Following this random motion, the atom may end up in the low momentum region,
where it stays for a long time, leading to the accumulation of atoms
mentioned above.

The time interval $\tau_s$ between two jumps is called the sojourn time.
It depends on the momentum of the atoms, and it also fluctuates
from one realization of the process to the other,
since photon scatterings occur in a random way.
An important notion to characterize the dynamics is the
mean sojourn time at momentum $p$, that is the
average time spent at momentum $p$ between two successive photon
scatterings.
This mean sojourn time is denoted as the lifetime $\tau(p)$,
and it takes a well-defined value for each momentum $p$.
The lifetime $\tau(p)$ is the inverse of the photon scattering rate
$R(p)$, so that the determination of the lifetime requires some knowledge
about the scattering rate.
Subrecoil laser cooling methods are based on atom-light interaction
processes such that the photon scattering rate $R(p)$ vanishes for $p = 0$.
In most cases,
$R(p)$ is a regular function of $p^2=\mathbf{p}^2$ due to rotational
invariance, so that the small $p$ behavior of $R(p)$ is given by
\be
R(p)= \tau_0^{-1}\, \left(\frac{p}{p_m}\right)^{\beta}, \qquad p \to 0,
\ee
where $\tau_0$ is a time scale, and $p_m$ is a momentum scale, to
be specified below.
The exponent $\beta$ is an even integer, $\beta=2$, $4$, or $6$,
which depends on the specific physical mechanisms used in the
experiment.\cite{BBE94,BBA02} It then follows that $\tau(p)$ behaves
at small $p$ as a (diverging) power law,
\be
\tau(p) = \tau_0 \left( \frac{p_m}{p} \right)^{\beta}, \qquad p \to 0.
\label{taup}
\ee
Hence for small enough $p$, the time spent at momentum $p$ becomes
very large with respect to the time spent at large momentum.
To be more specific, one can find a momentum scale $p_m \ll \hbar k$
(where $\hbar k$ is the typical change in atomic momentum
when a photon is scattered)
such that $R(p)$ behaves as a power law for $p<p_m$, and such that
the time spent by the atom in the region $p>p_m$ is very small
as compared to the time spent at $p<p_m$.
Note that, as $p_m \ll \hbar k$, values of $p$ smaller than $p_m$ are reached
in an essentially uniform way, after a photon scattering from a larger
momentum $p \approx \hbar k$.

Accordingly, to model in a simplified way subrecoil laser cooling, one
neglects the time spent outside the region $p<p_m$, and simply assumes that
the atom leaving this low-momentum region instantaneously comes back to it,
choosing at random a new momentum within the sphere
of radius $p_m$, with a uniform distribution in $d$ dimensions.
Thus the distribution $\rho(p)$ of the momentum $p$ reached after a
jump into the low-momentum region is
\be
\rho(p) = \frac{d}{p_m^d}\, p^{d-1}, \qquad 0<p<p_m.
\label{rhop}
\ee
Note that the absence of memory between two successive values of the
momentum means that the dynamics chosen here corresponds to a Markov process
--see Sect.~\ref{sect-lifetimes}.

Let us now introduce another probability distribution, namely the
probability $P_{st}(p)$ for an atom to have a momentum $p$ at a given time,
if a steady state is reached.
This probability is proportional to the lifetime $\tau(p)$, times the
probability $\rho(p)$ to reach a momentum $p$ after a jump.
Hence, $P_{st}(p)$ reads
\be
P_{st}(p) = \frac{1}{\langle \tau \rangle_p}\, \rho(p) \tau(p)
= \frac{\tau_0 p_m^{\beta}}{\langle \tau \rangle_p}
\, p^{-\beta} \rho(p),
\label{Pst}
\ee
where the average lifetime $\langle \tau \rangle_p$, computed over the
distribution $\rho(p)$, is given by
\be
\langle \tau \rangle_p \equiv \int_0^{p_m} \diff p \ \rho(p) \tau(p)
= d\tau_0 p_m^{\beta-d} \int_0^{p_m} \diff p \ p^{d-1-\beta}.
\label{Zp}
\ee
The integral in Eq.~(\ref{Zp}) converges only if $d>\beta$.
For $d \le \beta$, the average lifetime is infinite,
so that the distribution $P_{st}(p)$ given in Eq.~(\ref{Pst}) is not
normalizable.
Thus no stationary state can be reached, and the system continues to cool
for infinitely long times. 
Accordingly, the divergence of the average lifetime turns out to
play a key role in the phenomenon of subrecoil laser cooling,
as we shall see in more details in Sect.~\ref{sect-life-time}.
Note also that the divergence of $\langle \tau \rangle_p$ explains why
direct treatments of subrecoil laser cooling
using the equations of quantum optics become very difficult.\cite{BBA02}

\subsection{Aging and trap dynamics}

Glasses are materials for which the relaxation time becomes, at low
temperature, much larger than the accessible experimental time scales.
The phenomenology of glasses has been shown to be very rich, including in
particular non-exponential relaxations and aging
phenomena,\cite{Struik,Review}
in which case the relaxation time of the system
is proportional to its age (i.e., the time spent in the
low temperature phase).

\begin{figure}[t]
\centering\includegraphics[width=7.5cm,clip]{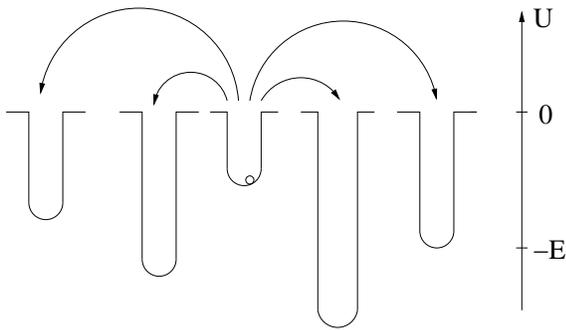}
\caption{\sl Schematic view of the energy landscape in the trap model:
a particle is trapped within deep energy wells, or traps, of energy $U=-E$
($E$ is called the depth of the trap) and can only escape through thermal
activation. Once the particle leaves the current trap, it chooses another
one at random among all the others.
}
\label{fig-trap}
\end{figure}

A simple model for glassy dynamics, that gives interesting insights
into the underlying physical mechanisms, is the trap model.
In this model, the system is represented by a particle in a high
dimensional phase space and is trapped into low energy states (traps),
as schematically shown in Fig.~\ref{fig-trap}.
These traps are characterized by their depth $E$
(i.e., the energy barrier to be overcome so as to leave the trap),
considered to be random variables with distribution $\rho(E)$.
The particle can leave a trap by thermal activation only,
and once a trap is left,
a new one is chosen at random, with uniform probability.
In physical terms, this means that the time spent by the system in high
energy states, that are not explicitely described in the model, is very
short.
In the simplest case, the distribution $\rho(E)$ is taken to be exponential
(which can be justified on the basis of
extreme value statistics\cite{BouchMez}),
\be
\rho(E) = \frac{1}{k_B T_g} \, e^{-E/k_B T_g}, \qquad E>0.
\label{rhoE}
\ee
where $k_B$ is the Boltzmann constant.
Up to now, $k_B T_g$ is simply an energy scale.
In the following, we consider the limiting case of an infinite number of
traps.

The dynamics of the model is defined by giving the distribution
$\varphi(\tau_s|E)$ of the sojourn times $\tau_s$ in a trap with depth $E$.
We consider here a Markov process, which means that sojourn times associated
to successively visited traps are uncorrelated.
For Markov processes, the distribution of sojourn times is exponential,
as discussed in Sect.~\ref{sect-lifetimes}.
We thus consider the following exponential form for 
$\varphi(\tau_s|E)$, namely
\be
\varphi(\tau_s|E) = \frac{1}{\tau(E)} \, e^{-\tau_s/\tau(E)},
\ee
where $\tau(E)$ is the mean sojourn time, i.e., the trap lifetime.
Note that the probability per unit time to escape
the trap is constant in a given trap, and is equal to $1/\tau(E)$.

From thermal activation arguments,
the trap lifetime $\tau(E)$, is given by the usual Arrhenius law,
\be
\tau(E) = \tau_0 \, e^{E/k_B T},
\label{arrhenius}
\ee
$\tau_0$ being a microscopic time scale that characterizes the random
motion of the particle within the trap.

In steady state, the probability to be in a trap of depth $E$ is
proportional to the trap lifetime, times the density distribution $\rho(E)$
of traps of depth $E$, leading to
\be
P_{eq}(E) = \frac{1}{\langle \tau \rangle_E}\, \rho(E)\, \tau(E)
= \frac{\tau_0}{\langle \tau \rangle_E} \, \rho(E)\, e^{E/k_B T}
\ee
which is nothing but the usual Boltzmann-Gibbs distribution (note the $+$
sign in the exponential, as $E$ is the opposite of the energy $U$ of the trap).
The average lifetime $\langle \tau \rangle_E$ is given by
\be
\langle \tau \rangle_E \equiv \tau_0 \int_0^{\infty} \diff E \ \rho(E)\,
e^{E/k_B T},
\ee
that is,
\be
\langle \tau \rangle_E = \frac{\tau_0}{k_B T_g} \int_0^{\infty}
\diff E \ \exp \left [\left( \frac{1}{T}-\frac{1}{T_g} \right)
\frac{E}{k_B} \right].
\ee
If $T \le T_g$, the equilibrium distribution $P_{eq}(E)$ ceases to be
normalizable since the integral defining $\langle \tau \rangle_E$ diverges.
Thus here again, the divergence of the average lifetime turns out to
play a major role, ruling out the existence of a steady state and leading
to a non-stationary dynamics at large times.
Accordingly, starting from a uniform distribution among traps
(to be interpreted as an infinite temperature initial condition), the average
dynamical energy of the system slowly drifts toward lower values,
without ever being able to reach a steady state. This will appear
explicitly in the next section.

\section{Unified description in terms of lifetimes} \label{sect-life-time}

\subsection{Distribution of lifetimes}
\label{sect-lifetimes}

We have seen that broad distributions of lifetimes can be found
in different physical problems belonging to different fields of physics.
These problems are usually studied using the natural physical variable,
like momentum for laser cooling,\cite{BBA02}
or energy barrier for the trap model.\cite{Monthus,BertinKov}
Yet, once expressed in terms of lifetimes, these
problems become equivalent from a formal point of view, at least when
considering the dynamical distribution of lifetimes.
Differences only appear when one wants to look at the predictions for the
physical variables, as discussed in Sect.~\ref{sect-link}.

Let us compute the distributions of lifetimes $\psi(\tau)$
for the laser cooling model as well as for the trap model.
Starting from Eqs.~(\ref{taup}) and (\ref{rhop})
--resp.~from Eqs.~(\ref{rhoE}) and (\ref{arrhenius})--
and using the relation $\rho(p) |dp| = \psi(\tau) |d\tau|$
--resp.~$\rho(E) |dE| = \psi(\tau) |d\tau|$--
one finds in both models a power law behavior for $\psi(\tau)$,
\be
\psi(\tau) = \frac{A}{\tau^{1+\alpha}},
\qquad \tau > \tau_{\rm min},
\ee
with $\alpha=d/\beta$ for laser cooling, and $\alpha=T/T_g$ for the trap
model; $A$ and $\tau_{\rm min}$ are model-dependent parameters, the precise
value of which being unimportant to the following discussion.
For $\alpha \le 1$, the first moment $\langle \tau \rangle$ diverges,
which means that the lifetime averaged over all possible states is formally
infinite. This has deep consequences on the dynamics, as discussed below,
since the system is in this case no longer able to reach a steady state.

At this stage, let us emphasize that two different notions
of time intervals appear in these problems,
namely the sojourn time $\tau_s$ and the lifetime $\tau$.
The sojourn time is the time spent between two successive events --scatterings
by a photon for laser cooling, or escapes from a trap in glassy dynamics.
The lifetime is associated to a microscopic `state' of the system,
and is defined as the average value of the sojourn time in this state.
Lifetimes can also be used to label the corresponding states,
instead of using the natural physical variables of the problem.
In addition, it is important to notice that the models considered here
are Markov processes.
This means that for a given lifetime $\tau$, the probability to exit the
current state during a short time interval $dt$ is $dt/\tau$.
As a result, the distribution $\varphi(\tau_s|\tau)$ of
sojourn times $\tau_s$ in a state with lifetime $\tau$ is exponential
for Markov processes,
\be
\varphi(\tau_s|\tau) = \frac{1}{\tau} \, e^{-\tau_s/\tau}.
\ee
Then the full distribution $\psi_s(\tau_s)$ of sojourn times,
averaged over all microscopic states, is
\be
\psi_s(\tau_s) = \int_0^{\infty} \diff \tau \ \varphi(\tau_s|\tau)
\, \psi(\tau)
= \int_0^{\infty} \frac{\diff \tau}{\tau} \ \psi(\tau)
\, e^{-\tau_s/\tau}.
\label{eq-taus-tau}
\ee
Note that if $\psi(\tau)$ has a power law tail for $\tau \to \infty$,
then $\psi_s(\tau_s)$ also has a power law tail, with the same exponent
as $\psi(\tau)$.

\subsection{Master equation for the dynamical distribution of lifetimes}

To obtain a unified picture for the physical models presented in
Sect.~\ref{sect-models}, we characterize the 
states of the system by their lifetimes $\tau$, without any reference to the 
physical meaning of these states. We wish to study the probability density 
$P(\tau,t)$ to be in a state with lifetime $\tau$ at time $t$.
For a Markov process,
the evolution of $P(\tau,t)$ is given by the following master equation
(see Appendix~\ref{apME}):
\bea \label{eqME0}
\frac{\partial P}{\partial t}(\tau,t) &=& -P(\tau,t) \int_0^{\infty} 
\diff \tau' \ W(\tau \to \tau')\\ \nonumber
&& \quad + \int_0^{\infty} \diff \tau'
\ W(\tau' \to \tau) P(\tau',t),
\eea
where $W(\tau \to \tau')$ is the transition rate from states with lifetime 
$\tau$ to states with lifetime $\tau'$.
The master equation is simply a balance equation for the probability
$P(\tau,t)$, taking into account the probability to leave states with
lifetime $\tau$, and the probability to reach them starting from
another state.
Consistently with the models presented in Sect.~\ref{sect-models},
we make the important assumption that 
the lifetimes $\tau_1$, $\tau_2$, ... $\tau_n$ of the states successively
occupied by the system are statistically independent.
Hence, at each transition, a new lifetime $\tau'$ is chosen at random
according to the distribution $\psi(\tau')$.
Given that the probability per unit time to leave state $\tau$ is $1/\tau$,
the transition rate thus takes the form
\be
W(\tau \to \tau') = \frac{1}{\tau} \psi(\tau').
\label{eqWUF}
\ee
The lack of correlation between the states before and after a jump
occurs physically because of the strongly stochastic underlying dynamics.
In subrecoil laser cooling, it comes from the fact that, in between two
trapping events, the atoms actually perform a random walk outside the trap,
which takes a negligible time, but uncorrelates two consecutive trapping
momenta.
A similar scenario holds for aging dynamics, where the system wanders
among high energy states
and fully decorrelates before finding a new trap.

Note that the independence property, which is stronger than the Markovian
property, would lead to a trivial stationary process if all states had
essentially the same lifetime.
However, as shown below, interesting non-stationary effects appear 
when the lifetimes are broadly distributed.

Using the form of the transition rates given in Eq.~(\ref{eqWUF}) 
and the normalization of $\psi(\tau')$, 
the master equation Eq.~(\ref{eqME0}) becomes
\be
\frac{\partial P}{\partial t}(\tau,t) = - \frac{1}{\tau} P(\tau,t) +
\psi(\tau) \int_0^{\infty} \frac{d\tau'}{\tau'} P(\tau',t).
\label{eqME1}
\ee

\subsection{Solution of the master equation}
\label{s3.C}

The master equation Eq.~(\ref{eqME1}) can be solved exactly using a Laplace
Transform with respect to time $t$.
Calculations are described in details in Appendix~\ref{apA}, and we
summarize the main results in the following.
The Laplace Transform $\hat{P}(\tau,s)$ of $P(\tau,t)$ is defined as
\be \label{laplace}
\hat{P}(\tau,s) \equiv \int_0^{\infty} \diff t \ e^{-st} P(\tau,t).
\ee
If we assume that at time $t=0$, the system is distributed completely at
random over all possible states, then the initial
distribution $P(\tau,t=0)$ reduces to $\psi(\tau)$.
In this case, $\hat{P}(\tau,s)$ can be written in a simple form
\be
\hat{P}(\tau,s) = \frac{1}{s\tau_e^*(s)} \,
\frac{\tau \psi(\tau)}{1+s\tau},
\label{eqPtau-s1}
\ee
where the time scale $\tau_e^*(s)$, to be interpreted as an
effective average lifetime, is defined as
\be \label{tau_e}
\tau_e^*(s) \equiv \int_0^{\infty} \diff \tau \
\frac{\tau \psi(\tau)}{1+s\tau}.
\label{def-taues}
\ee
Note that $\hat{P}(\tau,s)$ is normalized to $1/s$ with respect to $\tau$.
It should be emphasized that Eq.~(\ref{eqPtau-s1}) is an exact result
which does not rely on some long time regime ($s \ll \tau_0^{-1}$)
approximation. The influence on the dynamics
of the finiteness or divergence of the average lifetime
$\langle \tau \rangle$ now appears
clearly, since $\langle \tau \rangle = \lim_{s \to 0} \tau_e^*(s)$.
If $\langle \tau \rangle$ is finite, the distribution $\hat{P}(\tau,s)$
converges for $s \to 0$ (to be understood physically as $t \to \infty$)
to the equilibrium distribution (up to the usual $1/s$ factor)
\be
\hat{P}(\tau,s) \approx \frac{1}{s \langle \tau \rangle} \, \tau
\psi(\tau), \qquad s \to 0.
\label{eqPtau-s-stat}
\ee
Thus in this case, the system reaches an equilibrium state 
at infinite time.
Note however that the convergence to this asymptotic state may be very slow,
typically as a power law with an exponent which becomes small when
$\langle \tau \rangle$ is large.

In contrast, if $\langle \tau \rangle$ is infinite, the effective
average lifetime $\tau_e^*(s)$ grows without bound when $s \to 0$,
and no stationary state is ever reached.
If the lifetime distribution $\psi(\tau)$ has a power law tail,
\be
\psi(\tau) \sim \frac{c}{\tau^{1+\alpha}}, \qquad \tau \to \infty,
\ee
with $0<\alpha<1$, then the out-of-equilibrium solution of the master
equation exhibits some interesting non trivial properties, in particular
a scaling form at long times, that is for $s \to 0$.
To be more specific, the distribution $\hat{P}(\tau,s)$ can be written
asymptotically as
\be
\hat{P}(\tau,s) \approx \tilde{\phi}(s\tau), \qquad s \to 0,
\label{eq-scal}
\ee
where the scaling function $\tilde{\phi}(x)$ can be determined from
Eq.~(\ref{eqPtau-s1}), namely
\be
\tilde{\phi}(x) = \frac{\pi}{\sin(\pi\alpha)} \, \frac{1}{x^{\alpha}(1+x)}.
\label{eq-phi}
\ee
To come back to physical time, the inverse Laplace Transform of
$\hat{P}(\tau,s)$ in the scaling regime can be computed. As expected,
the resulting distribution $P(\tau,t)$ also takes a scaling form,
\be
P(\tau,t) = \frac{1}{t} \, \phi \left( \frac{\tau}{t} \right),
\label{eqscal-tUF}
\ee
which shows that the typical momentum $\tau_t$
at time $t$ is of the order of $t$
itself. The scaling function $\phi(u)$ is such that
$\phi(u) \sim u^{-\alpha}$ for $u \to 0$ and $\phi(u) \sim u^{-1-\alpha}$
for $u \to \infty$
(note that $\tilde{\phi}$ is not the Laplace Transform of $\phi$).
The asymptotic behaviour of $P(\tau,t)$
for $\tau \ll t$ or $\tau \gg t$ can then be computed, and is found to be
qualitatively the same as that of $\hat{P}(\tau,s=1/t)$.
Neglecting multiplicative constants, the distribution
$P(\tau,t)$ takes for $\tau \ll t$ a pseudo-equilibrium form,
\be \label{tau-ll-t}
P(\tau,t) \sim \tau \, \psi(\tau), \qquad \tau \ll t,
\ee
whereas for $\tau \gg t$, it remains proportional to the
a priori distribution $\psi(\tau)$,
\be \label{tau-gg-t}
P(\tau,t) \sim \psi(\tau), \qquad \tau \gg t.
\ee
This means that states with small enough lifetimes have been visited
a large number of times and are essentially equilibrated. In contrast,
states with very large lifetimes have been visited only once, and the precise
value of their lifetime is not felt by the dynamics at this stage.

\subsection{Link between lifetime and physical observables}
\label{sect-link}

Even though the different models introduced in Sect.~\ref{sect-models} share a
common dynamical distribution $P(\tau,t)$,
physical observables do not behave in the same way in these models.
The reason for this is that the functional relations between these
observables and the lifetimes differ much from one model to the other.

In the subrecoil laser cooling problem, the relation
between momentum $p$ and lifetime $\tau$ is
$p = p_m (\tau/\tau_0)^{-1/\beta}$.
Hence from Eq.~(\ref{eqscal-tUF}), the dynamical distribution of momentum
$P_{\mathrm{m}}(p,t)$ can be written, in the long time regime, as
\be
P_{\mathrm{m}}(p,t) = \frac{\beta \tau_0 p_m^{\beta}}{t p^{1+\beta}} \,
\phi\left( \frac{\tau_0 p_m^{\beta}}{t p^{\beta}}\right).
\ee
Thus the distribution of momentum becomes more and more narrow,
and the typical momentum $p^*(t)$ reached after a large time $t$ is
given by
\be
p^*(t) \approx p_m (t/\tau_0)^{-1/\beta}.
\ee
which shows that the atoms can, in principle, be cooled down to
arbitrarily low temperatures, provided that one waits long enough, and that
no other effects are present. 
The above result is indeed in remarkable agreement with the observations made
in laser cooling experiments,\cite{BBA02} and leads to a clear
understanding of this intriguing phenomenon, of purely dynamical origin.

In the trap model of glass aging, the energy $E$ is related to the lifetime
$\tau$ through $E=-T\ln (\tau/\tau_0)$.
Since the typical lifetime $\tau_t$ at time $t$ is of the order of $t$,
the average energy $E^*(t)$ behaves as
\be
E^*(t) \approx -T \ln (t/\tau_0).
\label{dyn-energy}
\ee
Such a logarithmic behavior of the energy is indeed frequently observed
for aging systems, and is accounted for in a simple way in the present
approach. Indeed, the latter gives a precise meaning to the word 'aging',
as it is clear from the scaling form Eq.~(\ref{eqscal-tUF})
and from the average energy given in Eq.~(\ref{dyn-energy}) that the system
keeps track of its age.

\section{Conclusion}

In this note, we illustrated on two simple examples how the apparently
formal broad distributions can lead to deep insights and
to a unified view on the physics of systems that do not reach
an equilibrium state. This formalism leads in a straightforward way
to the long-time behavior of such systems, and shows that this behavior
is universal, only depending on the exponent $\alpha$ of the
power-law tail of the lifetime distribution.
In addition, this approach gives a well-defined meaning to the notion
of ``quasi-equilibrium'' at large times, which is often referred to
in such systems. In a quasi-equilibrated system at a large time $t$,
microscopic states with lifetimes less than $t$ are
essentially equilibrated,
whereas those with a lifetime greater than $t$ are not.
Finally, let us mention that more general situations, for which the
return time between two trapping events is not negligible,
or even broadly distributed
(while this time was neglected in the present approach),
may also be considered. Such a generalization does not change the main
qualitative conclusions reported here.\cite{BBA02}

\section{Acknowledgments}

E.B. is greatly indebted to Alain Aspect, Jean-Philippe Bouchaud and
Claude Cohen-Tannoudji (coauthors with F.B.
of a book on L\'evy statistics and laser cooling\cite{BBA02})
for their help in completing this work
after Fran\c{c}ois Bardou passed away.

\appendix

\section{Master equation formalism}
\label{apME}

On general grounds, let us consider a system described by a finite
set of configurations (or microscopic states) $\{\mathcal{C}\}$.
The set $\{\mathcal{C}\}$ may be called the configuration space of the system.
We assume that the system evolves according to a markovian stochastic
dynamics, which means that the system may change configuration randomly
in time, and without memory of the configurations occupied earlier.
Such a random dynamics is often called a Markov process.
To be more specific, let us assume that the system is in an arbitrary
configuration $\mathcal{C}$ at time $t$. The probability that the system
goes from $\mathcal{C}$ to a new configuration $\mathcal{C}'$ between time $t$ and
$t+dt$, where $dt$ is an infinitesimal time interval, is equal to
$W(\mathcal{C} \to \mathcal{C}') dt$, which defines the transition rate
(i.e., probability per unit time)
from $\mathcal{C}$ to $\mathcal{C}'$, $W(\mathcal{C} \to \mathcal{C}')$.
Roughly speaking, the markovian property of the dynamics states that
$W(\mathcal{C} \to \mathcal{C}')$ depends neither on the history of the system
before arriving at configuration $\mathcal{C}$, nor on the time spent
at configuration $\mathcal{C}$. Let us also emphasize that
$W(\mathcal{C} \to \mathcal{C}')$ should be understood as the probability
per unit time
to go to configuration $\mathcal{C}'$ \emph{given} that the system is in
configuration $\mathcal{C}$; it is thus a conditional probability.

Let us now introduce the probability $P_{\mathcal{C}}(t)$ for the system
to be in a configuration $\mathcal{C}$ at time $t$. The evolution with time
of $P_{\mathcal{C}}(t)$ is described by a master equation, which is nothing but
a balance equation for the probability. It may be obtained in the
following way. In a very short time interval $dt$, the variation
$dP_{\mathcal{C}}$ of $P_{\mathcal{C}}(t)$ is due on the one hand,
to the probability
$d\Phi_{\mathcal{C}}^{\rm out}$ to leave the configuration $\mathcal{C}$
during the time interval $dt$, and on the other hand to the probability
$d\Phi_{\mathcal{C}}^{\rm in}$
to reach the configuration $\mathcal{C}$ from any other configuration
$\mathcal{C}'$ in the same time interval. This yields
\be \label{general-ME}
dP_{\mathcal{C}} = - d\Phi_{\mathcal{C}}^{\rm out} +
d\Phi_{\mathcal{C}}^{\rm in}.
\ee
These two contributions can be evaluted as follows.
Starting from $\mathcal{C}$,
the system may go to any other configuration $\mathcal{C}'$, so that the
corresponding probability is simply the sum of all such possible transition
probabilities. The probability to be in configuration $\mathcal{C}$ and to
go to configuration $\mathcal{C}'$ within $dt$ is $W(\mathcal{C} \to
\mathcal{C}') P_{\mathcal{C}}(t) dt$, leading to
\be
d\Phi_{\mathcal{C}}^{\rm out} = \sum_{\mathcal{C}' \ne \mathcal{C}}
W(\mathcal{C} \to \mathcal{C}') P_{\mathcal{C}}(t) dt.
\ee
In a similar way, $d\Phi_{\mathcal{C}}^{\rm in}$ is obtained by summing
the transition probabilities from all other configurations $\mathcal{C}'$
to configuration $\mathcal{C}$,
\be
d\Phi_{\mathcal{C}}^{\rm in} = \sum_{\mathcal{C}' \ne \mathcal{C}}
W(\mathcal{C}' \to \mathcal{C}) P_{\mathcal{C}'}(t) dt.
\ee
Substituting into Eq.~(\ref{general-ME}), one obtains
\be
\frac{dP_{\mathcal{C}}}{dt} = - \sum_{\mathcal{C}' \ne \mathcal{C}}
W(\mathcal{C} \to \mathcal{C}')
P_{\mathcal{C}}(t) + \sum_{\mathcal{C}' \ne \mathcal{C}} W(\mathcal{C}'
\to \mathcal{C}) P_{\mathcal{C}'}(t),
\ee
which is precisely the master equation describing the stochastic evolution
of the system. Note that for simplicity, the above argument was restricted to
systems with a finite number of discrete configurations.
This formalism may be extended to describe systems with
continuous configuration space, as done in the present note.

\section{Explicit solution of the master equation}
\label{apA}

We report in this appendix the exact solution, in Laplace transform, of the
master equation given in Eq.~({\ref{eqME1}}).
Introducing $\hat{P}(\tau,s)$ defined in Eq.~(\ref{laplace}), 
the master equation becomes:
\be
s \hat{P}(\tau,s) - P_0(\tau) = -\frac{1}{\tau} \hat{P}(\tau,s) + \psi(\tau)
\int_0^{\infty} \frac{\diff \tau'}{\tau'} \ \hat{P}(\tau',s),
\ee
where $P_0(\tau)$ is the initial condition, $P_0(\tau) \equiv P(\tau,t=0)$. 
Solving for $\hat{P}(\tau,s)$, one finds
\be
\hat{P}(\tau,s) = \frac{\tau P_0(\tau)}{1+s\tau} +
\frac{\tau \psi(\tau)}{1+s\tau} \, \hat{\omega}(s),
\label{eqMElap1}
\ee
with $\hat{\omega}(s) = \int_0^{\infty} \diff \tau' \, \tau'^{-1}
\ \hat{P}(\tau',s)$.
One first needs to solve for $\hat{\omega}(s)$. Dividing
Eq.~(\ref{eqMElap1}) by $\tau$ and integrating over $\tau$, one has
\be
\int_0^{\infty} \frac{\diff \tau}{\tau} \ \hat{P}(\tau,s)
= \int_0^{\infty} \diff \tau \ 
\frac{P_0(\tau)}{1+s\tau} + \hat{\omega}(s) \int_0^{\infty} \diff \tau \ 
\frac{\psi(\tau)}{1+s\tau}.
\ee
Solving for $\hat{\omega}(s)$ yields
\be
\hat{\omega}(s) = \frac{1}{s} 
\left( \int_0^{\infty} \diff \tau \ \frac{\tau \psi(\tau)}{1+s\tau}
\right)^{-1} \int_0^{\infty} \diff \tau \ \frac{P_0(\tau)}{1+s\tau}.
\ee
Replacing in Eq.~(\ref{eqMElap1}), one gets
\bea
\hat{P}(\tau,s) &=& \frac{\tau P_0(\tau)}{1+s\tau} + \frac{1}{s} \,
\frac{\tau \psi(\tau)}{1+s\tau} \\ \nonumber
&\times& \left( \int_0^{\infty} \diff \tau' \ 
\frac{\tau' \psi(\tau')}{1+s\tau'} \right)^{-1}
\int_0^{\infty} \diff \tau' \ \frac{P_0(\tau')}{1+s\tau'}.
\eea
Using the time scale $\tau_e^*(s)$ introduced in
Eq.~(\ref{tau_e}), the solution of the master equation can be written as
\be
\hat{P}(\tau,s) = \frac{\tau P_0(\tau)}{1+s\tau} + \frac{1}{s \,
\tau_e^*(s)} \, \frac{\tau \psi(\tau)}{1+s\tau}
\int_0^{\infty} \diff \tau' \ \frac{P_0(\tau')}{1+s\tau'}.
\label{eqPtau-s0}
\ee
The quantity $\tau_e^*(s)$ may be interpreted
as an effective average lifetime,
in the sense that it is defined as an average of $\tau$ over the distribution
$\psi(\tau)$, taking into account a cut-off of dynamical origin around
a time scale $t^* \approx s^{-1}$.
In the specific case where $P_0(\tau)=\psi(\tau)$, 
Eq.~(\ref{eqPtau-s0}) simplifies to Eq.~(\ref{eqPtau-s1})
after a straightforward calculation.

Note that it is interesting to introduce a second time scale 
$\tau_0^*(s)$ associated to $P_0(\tau)$ through
\be
\tau_0^*(s) \equiv \int_0^\infty \diff \tau \ 
\frac{\tau P_0(\tau)}{1 + s\tau},
\label{eq-tau0s}
\ee
since one can then rewrite Eq.~(\ref{eqPtau-s0}) in the more compact form
\bea \label{eqPtau-s2}
\hat{P}(\tau,s) &=& \frac{1}{s\tau_e^*(s)} \, \frac{\tau}{1+s\tau}\\
\nonumber
&& \quad \times 
\left[ s\tau_e^*(s) \, P_0(\tau) + \left( 1-s\tau_0^*(s)
\right) \, \psi(\tau) \right].
\eea
The time scale $\tau_0^*(s)$ is the effective average time
associated with the initial distribution $P_0(\tau)$, taking again into
account the cut-off time scale $t^* \approx s^{-1}$.

In the large time limit, only the asymptotic behavior of $\hat{P}(\tau,s)$
for $s \to 0$ remains relevant.
In this limit, the expression between square brackets in Eq.~(\ref{eqPtau-s2})
reduces to $\psi(\tau)$, as $s\tau_e^*(s)$ and
$s\tau_0^*(s)$ go to zero
(even though $\tau_e^*(s)$ and $\tau_0^*(s)$
may diverge when $s \to 0$), and one recovers the same result as
Eq.~(\ref{eqPtau-s1}): the memory of the initial condition is lost.
In particular, the scaling form Eq.~(\ref{eq-scal}) and the functional
form of the scaling function $\tilde{\phi}(x)$ given in Eq.~(\ref{eq-phi})
do not depend on the initial condition.
Yet, the crossover time $t \approx s^{-1}$
beyond which the scaling form Eq.~(\ref{eq-scal}) 
becomes valid depends on $P_0(\tau)$. For instance, if $P_0(\tau)$
has a peak around a value $\tau_{\mathrm{pk}} \gg \tau_0$, the
scaling regime is reached only for times $t \gg \tau_{\mathrm{pk}}$,
so that it may be
necessary to wait for an arbitrarily large time.

Performing the inverse Laplace transform of Eq.~(\ref{eq-scal})
then leads to the scaling form (\ref{eqscal-tUF}), where the
scaling function $\phi(u)$ is given by\cite{Monthus}
\be
\phi(u) = \frac{\sin \pi \alpha}{\pi \Gamma(\alpha)} \, \frac{1}{u}
\, e^{-1/u} \int_0^{1/u} \diff v \ v^{\alpha-1} e^v.
\label{eq-phi-u}
\ee
An asymptotic expansion of Eq.~(\ref{eq-phi-u})
for large and small values of $u$ respectively
leads to the asymptotic behavior of $P(\tau,t)$ given in
Eqs.~(\ref{tau-ll-t}) and (\ref{tau-gg-t}).

\section{Suggested project}
\label{sugg-pb}

The dynamical distribution $P(\tau,t)$ and the scaling behaviour
given in Eq.~(\ref{eqscal-tUF}) can be obtained using a simple
numerical simulation of the random process.
Drawing a random number $v$ from a uniform distribution
between $0$ and $1$, compute the lifetime $\tau$ of
the current state as $\tau=\tau_0 \, v^{-1/\alpha}$
(the microscopic time $\tau_0$ may be set to unity).
Check that the probability distribution $p(\tau)$ satisfies
\be
p(\tau) = \frac{\alpha\, \tau_0^{\alpha}}{\tau^{1+\alpha}},
\qquad \tau > \tau_0.
\ee
This can be done either numerically, or analytically noticing that
$p(\tau) |\diff \tau|=|\diff v|$.
Once a state with lifetime $\tau$ is given, the sojourn time $\tau_s$
in this state is chosen randomly from an exponential distribution of
mean $\tau$. To do this, draw a uniform random number $u$, $0<u<1$,
and compute $\tau_s=-\tau \ln u$. Check again that $\tau_s$ has the
desired distribution. Then iterate this computation and sum
the different sojourn times until the final time $t$ is reached
($t$ is an arbitrary fixed time such that $t \gg \tau_0$,
typically in the range $t=10^3 \,\tau_0$ to $10^6 \,\tau_0$),
and record the lifetime $\tau$ of the state occupied at time $t$.

Repeat this procedure a large number
of times, and compute the histogram of the recorded values of $\tau$;
normalizing the histogram yields the distribution $P(\tau,t)$.
Compute this distribution for different times, and check that for
$\alpha>1$, it becomes independent of $t$ in the large $t$ limit,
whereas for $\alpha<1$, the $t$-dependence remains for arbitrarily large $t$.
In the latter case, check the scaling form Eq.~(\ref{eqscal-tUF}), and
the asymptotic behavior of $P(\tau,t)$ for both $\tau \ll t$
and $\tau \gg t$.

Alternatively, these simulations may be carried out by considering
the momentum $p$ or the energy $E$ instead of the lifetime $\tau$.

\end{document}